\documentclass[10pt,conference]{IEEEtran}

\usepackage[utf8]{inputenc}
\usepackage{url}
\usepackage{colortbl}
\usepackage{balance}
\usepackage{xspace}
\usepackage{graphicx}
\usepackage{verbatim}
\usepackage{bold-extra}
\usepackage{amsmath}
\usepackage{multirow}
\usepackage{listings}
\usepackage{boxedminipage}
\usepackage{soul}
\usepackage{enumitem}
\usepackage[normalem]{ulem}
\setlist{nolistsep,leftmargin=.5cm}
\usepackage{booktabs}
\usepackage{tabularx}
\usepackage{svg}
\usepackage{calc}
\usepackage{float}
\usepackage{multirow}
\usepackage[normalem]{ulem}
\useunder{\uline}{\ul}{}
\usepackage{amsmath}
\usepackage[most]{tcolorbox}
\definecolor{MidnightBlue}{HTML}{006895}
\definecolor{BoxesBlue}{HTML}{DEECFF}
\definecolor{BoxesYellow}{HTML}{FFF2CC}
\definecolor{StateGreen}{HTML}{91C788}
\definecolor{StateRed}{HTML}{FF8080}
\definecolor{ArrowGreen}{HTML}{61B15A}
\definecolor{ArrowViolet}{HTML}{BA94D1}
\usepackage{caption}
\usepackage{microtype} 
\usepackage[noadjust]{cite}
\usepackage{amsmath,amssymb,amsfonts}
\usepackage{algorithmic}
\usepackage{graphicx}
\usepackage{textcomp}
\usepackage{xcolor}
\usepackage[hidelinks]{hyperref}
\usepackage{ifthen}
\usepackage{wrapfig}
\usepackage{subcaption}
\usepackage{cleveref} 
\usepackage{enumitem}
\usepackage{indentfirst}

\newboolean{showcomments}
\setboolean{showcomments}{false}

\newboolean{showboxes}
\setboolean{showboxes}{true}

\usepackage{adjustbox}
\usepackage{totcount}

\ifthenelse{\boolean{showcomments}}

\setboolean{showcomments}{true}

\ifthenelse{\boolean{showcomments}}
{\newcommand{\nb}[2]{
		\fbox{\bfseries\sffamily\scriptsize#1}
		{\sf\small$\blacktriangleright$\textit{#2}$\blacktriangleleft$}
	}
}
{\newcommand{\nb}[2]{}
}

\newcommand\rev[1]{{\color{black}{#1}}}

\newcommand{\ie}{\textit{i.e.},\xspace}
\newcommand{\eg}{\textit{e.g.},\xspace}

\newcommand{\etal}{\textit{et al.}\xspace}
\newcommand{\aka}{\textit{a.k.a.}\xspace}
\newcommand{\toolname}{\textsc{Sprint}\xspace}

\usepackage{tikz}

\definecolor{bug_red}{rgb}{.84,.23,.29}

\definecolor{info-needed-color}{rgb}{1,.8,.12}

\newcounter{findingcounter}
\newtotcounter{totalfindings}
\ifthenelse{\boolean{showboxes}}
{
    \newcommand{\finding}[1]{%
      \refstepcounter{findingcounter}
      \begin{tcolorbox}[boxsep=1pt,left=2pt,right=2pt,top=1pt,bottom=1pt]%
      \small
      \centering
      \textbf{Finding \arabic{findingcounter}:} #1
      \end{tcolorbox}%
      \addtocounter{totalfindings}{1}
    }
    
  }
{
    \newcommand{\finding}[1]{}
}

\ifthenelse{\boolean{showboxes}}
{
	\newcommand{\rqanswer}[1]{%
		\begin{tcolorbox}[enhanced,skin=enhancedmiddle,borderline={1mm}{0mm}{MidnightBlue},boxsep=3pt]
			\small
			#1
		\end{tcolorbox} 
    }
}
{
	\newcommand{\rqanswer}[1]{}
}

\begin{document}

\title{\toolname: An Assistant for Issue Report Management
}

\author{
\IEEEauthorblockN{Ahmed Adnan}
\IEEEauthorblockA{
\textit{University of Dhaka}\\
Dhaka, Bangladesh \\
\href{mailto:}{bsse1131@iit.du.ac.bd}}
\and
\IEEEauthorblockN{Antu Saha}
\IEEEauthorblockA{
\textit{William \& Mary}\\
Williamsburg, Virginia, USA \\
\href{mailto:}{asaha02@wm.edu}}
\and
\IEEEauthorblockN{Oscar Chaparro}
\IEEEauthorblockA{
\textit{William \& Mary}\\
Williamsburg, Virginia, USA \\
\href{mailto:}{oscarch@wm.edu}}
}


%
%

\maketitle

\begin{abstract}
Managing issue reports is essential for the evolution and maintenance of software systems. However, manual issue management tasks such as triaging, prioritizing, localizing, and resolving issues are highly resource-intensive for projects with large codebases and users. To address this challenge, we present \toolname, a GitHub application that utilizes state-of-the-art deep learning techniques to streamline issue management tasks. \toolname assists developers by: (i) identifying existing issues similar to newly reported ones, (ii) predicting issue severity, and (iii) suggesting code files that likely require modification to solve the issues. We evaluated \toolname using existing datasets and methodologies, measuring its predictive performance, and conducted a user study with five professional developers to assess its usability and usefulness. The results show that \toolname is accurate, usable, and useful, providing evidence of its effectiveness in assisting developers in managing issue reports. \toolname is an open-source tool available at \href{https://github.com/sea-lab-wm/sprint_issue_report_assistant_tool/}{github.com/sea-lab-wm/sprint\_issue\_report\_assistant\_tool}.
\end{abstract}



\section{Introduction}
\label{sec:intro}

Issue reports are essential artifacts to identify, track, and resolve issues in software systems~\cite{blImportance, detmisinginfoinbr, bettenburg2008makes,Saha:icse25}. Given their importance, managing issue reports is critical for maintaining and evolving software systems~\cite{benefitsofissuemanage, brAnalysis}. Key issue management activities include labeling issues that report similar problems, categorizing them based on severity, and identifying potential buggy locations in the code~\cite{dupbrharmful, otoom2019automated, blImportance}. While important, manual issue management is a time-consuming and challenging process, especially for projects that typically receive hundreds of issues affecting various software components.  

To help developers manage issues, researchers have proposed a variety of techniques to automate issue report management tasks~\cite{practitionerPerceiveAutomation}.  Practitioners have also proposed a variety of tools~\cite{find_duplicates,probot,rocha2015nextbug,priorityScheduler,pr-agent}, standalone or integrated into existing systems (\eg issue trackers), to support these tasks. However, most of these tools are designed to support specific tasks and their underlying models are not easy to update. As such, these tools do not integrate different techniques and support for various issue management tasks in a single, easy-to-extend, comprehensive solution.
\looseness=-1



To address these limitations, we introduce \toolname, an integrated GitHub application for issue management. \toolname is designed as a comprehensive solution that can be easily extended and adapted to support multiple issue management tasks. In its current version, \toolname leverages state-of-the-art deep learning techniques to assist developers in (i) identifying issues similar to the newly reported issue, (ii) predicting issue severity, and (iii) localizing the potential code files that require modification to solve the reported problem. \toolname gives the developers suggestions as comments in issue reports and as labels attached to the issues to facilitate issue management. \toolname is extensible due to its modularized plugin-based architecture, which includes well-designed APIs and scripts that enable easy feature integration and extensibility. \toolname is also scalable, allowing multiple users and repositories to leverage its features concurrently. 

We evaluated the predictive performance of \toolname's underlying state-of-the-art models by replicating the evaluations of their respective papers. Additionally, we conducted a user study with five professional developers experienced in issue management, who found \toolname easy to use, useful, and practical for issue management and resolution. \toolname is an open-source tool hosted on GitHub~\cite{repl_pack} and can be easily installed in any GitHub repository.


\section{\toolname: An Issue Report Management Tool}
\label{sec:tool}

\subsection{Supported Issue Management Tasks}
\toolname is an issue management assistant for developers, project managers, computer science students, and educators. \toolname currently supports three issue management tasks.

\subsubsection{Issue Severity Prediction}
\toolname classifies the reported issues based on their severity level. After a user creates a new issue, the tool tags it with one of five labels: Blocker, Critical, Major, Minor, or Trivial. This helps project managers and developers prioritize the issues that require immediate attention.
\looseness=-1

\subsubsection{Similar Issue Identification}
\toolname suggests similar issues as soon as a new issue is submitted, tagging the new issue with a ``Duplicate” label. Similar issues are suggested in a comment in the issue report.  
This feature minimizes redundant issue management efforts by suggesting related issues to developers and reporters, who are meant to inspect the results to determine if the new issue was reported before.
\looseness=-1 

\subsubsection{Buggy Code Localization}
\toolname suggests potential buggy code files based on the textual similarity between the reported issue and the code file of the system's latest version. When a new issue is reported, \toolname fetches the files of the latest system version and ranks them as a list of potential buggy code files. This feature suggests users code files that might require modification to solve the issues.


\begin{figure}[t]
    \centering
    \includegraphics[width=0.95\columnwidth, keepaspectratio]{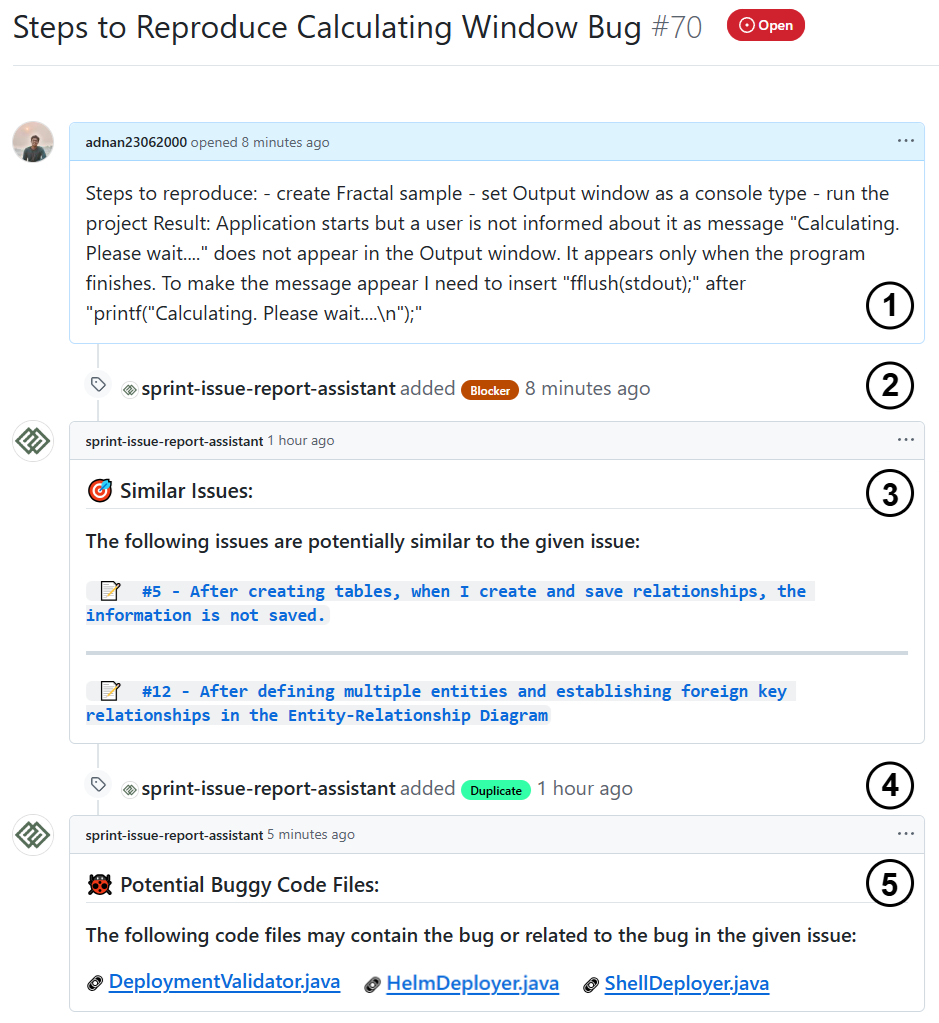}
    \caption{\toolname's GUI with a Usage Scenario}
    \label{fig:example-image}
\end{figure}

\subsection{Usage Scenario \& Graphical User Interface}

\toolname can be easily installed in any repository. A user only needs to visit the installation website~\cite{sprintUrl}, click the `Install' button, and select the repositories where the tool will be used without requiring any configuration. 
\looseness=-1

After installation, when a user reports a new issue  (see \textcircled{1} in \cref{fig:example-image}), \toolname's first step is to fetch the issue's title and description, and the code files of the system's latest version. \toolname then generates comments with feedback to the user by analyzing this information using state-of-the-art models (described in \cref{sec:architecture}). 

First, \toolname analyzes the new issue and classifies it into one of five severity levels (Blocker, Critical, Major, Minor, or Trivial~\cite{severityTypes}), assigning a label with predicted severity \textcircled{2}. This label follows a color-coding format from red to yellow, where red suggests the issue is very severe and yellow indicates the issue is trivial. 
Second, \toolname analyzes the new issue and all the existing issues in the project, suggesting which of these are similar to the new issue. Then, it generates a comment with a list of suggested similar issues \textcircled{3}, each including its issue ID, title, and URL. If one or more similar issues are found, the new issue is labeled as ``Duplicate" \textcircled{4}.  Third, \toolname identifies potential buggy code files by analyzing the reported issue's information and the paths and names of the repository's code files.
\toolname generates a comment displaying the list of files (with URLs) that may need modification to solve the issue \textcircled{5}.
\toolname's features are independent of one another: no feature is dependent on the execution of others.
\looseness=-1

\toolname suggests severity labels, similar issues, and potential buggy code files for the issues created after installation, whenever a new issue is submitted.  
\toolname does not generate comments or labels for the issues existing before installation because these suggestions might conflict with comments and labels manually created by developers and reporters. 
\toolname currently handles the code files of the system's latest version. A future tool improvement is to identify the affected system version specified in the issue and perform bug localization on that version's code files.
\section{\toolname’s Architecture \& Implementation}
\label{sec:architecture}

\subsection{Architecture}
\begin{figure}[t]
    \centering
    \includegraphics[width=0.5\textwidth]{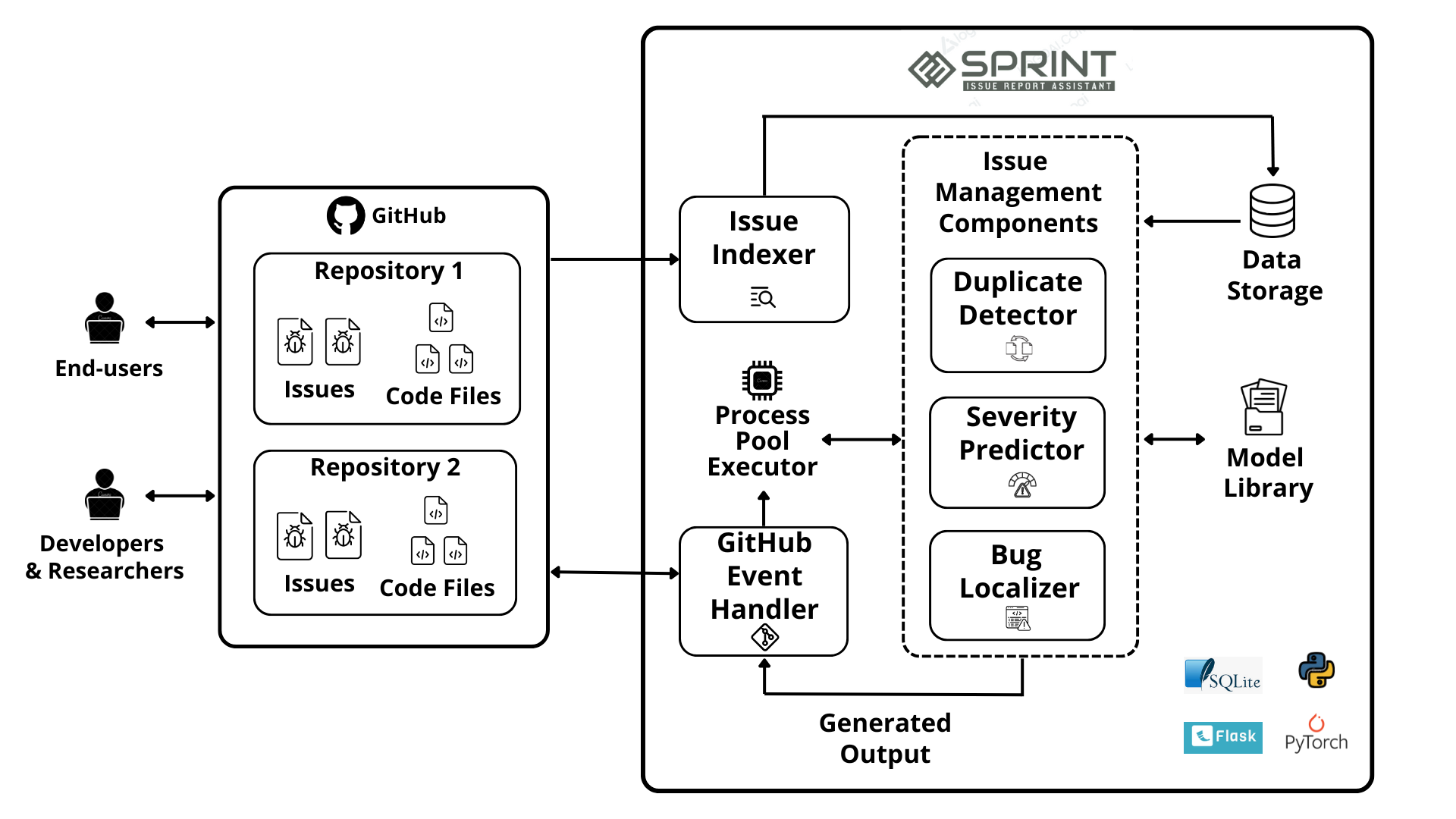} 
    \caption{Overview of \toolname's Architecture}
    \label{fig:architecture}
\end{figure}

\toolname's architecture, shown in \cref{fig:architecture}, consists of three main components: (1) the  Issue Indexer, (2) the GitHub Event Listener, and (3) the Issue Management Components.
\subsubsection{Issue Indexer}
When \toolname is installed in one or more repositories, this component fetches all the existing issues from those repositories using GitHub Webhooks~\cite{webhookDoc} and stores them in a local relational database. This component applies page-based indexing to partition issues into manageable groups for efficient fetching. The database is meant to facilitate quick access and analysis of issues. This database is permanently synchronized with GitHub to provide 24/7 support since new issues are reported continuously and concurrently.
\subsubsection{GitHub Event Handler}
This component is responsible for listening to and handling the GitHub repository events when a new issue is submitted. This component fetches a newly reported issue along with the latest version's code files and sends them to the Issue Management Components for further analysis. After the analysis, this component processes the feedback, formats it appropriately, and posts the generated comments and labels to the reported issue as the final output.

\subsubsection{Issue Management Components}
There are three components for issue management in the current version of \toolname.

\underline{Similar Issue Detection:} This component takes a newly created issue (title \& description) from the GitHub Event Handler and compares it with each issue stored in the database by analyzing textual similarity between them. For this task, \toolname uses the RTA classification model~\cite{fang2023representthemall}, fine-tuned on the RTA duplicate bug report training dataset~\cite{representThemAllDataset}. \toolname uses a Process Pool Executor~\cite{processPool}, a multiprocessing component, to analyze multiple issue pairs concurrently. After the analysis, this component returns the duplicate issues to the GitHub Event Handler.
\looseness=-1

\underline{Issue Severity Prediction:} This component receives a newly reported issue (issue title \& description) from the GitHub Event Handler and utilizes the RTA classification model~\cite{fang2023representthemall}, fine-tuned on RTA severity prediction training dataset~\cite{representThemAllDataset}, to classify the issue into one of five severity levels: blocker, critical, major, minor, and trivial. This component returns the predicted issue severity level to the Event Handler.
\rev{For similar issue detection and severity prediction tasks, we selected RTA~\cite{fang2023representthemall} because of its state-of-the-art predictive performance on large open-source issue reports and its speed and efficiency in issue analysis. It learns the fundamental representation of bug reports via a dynamic masked language model and contrastive learning objectives in a self-supervised manner.}

\underline{Bug Localization:} This component receives a newly created issue (title \& description) and the code files of the latest system version from the GitHub Event Handler. For bug localization, \toolname uses the bug localization approach proposed by Bogomolov \etal~\cite{longCodeArena}: a Llama-2-7b-chat~\cite{llama-2} model, fine-tuned on their bug localization dataset~\cite{longCodeBLDataset}. 
The bug localization component constructs a prompt including the issue contents and the list of code file paths and names. With this prompt, the model generates a ranked list of potential buggy code files that might need further inspection to solve the issue and returns it to the Event Handler. The prompt asks the model to compare the textual semantics between the issue content and repository code file paths/names to predict potential buggy code files.

\subsection{\toolname's Extensibility \& Scalability}
\toolname's architecture separates the issue management components from the components used for GitHub integration and event handling. The features are implemented as modular APIs with comprehensive documentation to simplify tool feature addition and enhancement. Our tool also conforms with the plugin architecture~\cite{pluginArchitecture} to enable easy extensibility. The event handler serves as the host component, while the issue management component APIs serve as the plugin interface. As per plugin architecture, \toolname's GitHub application supports dynamic loading and communication protocols, ensuring isolation between the host and plugins. Moreover, in the tool repository~\cite{repl_pack}, we provided scripts that can be used to fine-tune other transformer-based models for issue management tasks. With this, project owners can easily integrate the resulting models into our tool by replacing the model paths in the tool's configuration file.

\toolname's backend utilizes Python's Process Pool Executor, which applies multiprocessing to concurrently handle multiple issues from various repositories. Project owners can configure the pool size based on their computational resources and workload requirements, by changing the configuration file, allowing \toolname to be responsive even during simultaneous requests. This multiprocessing technique also facilitates duplicate detection by notably reducing processing time and ensuring fast responses since pairwise issue comparison is computationally intensive.
\looseness=-1


\subsection{Implementation}
\toolname is implemented using Python's Flask framework~\cite{pythonFlask}. \toolname is developed as a GitHub Application, using GitHub’s Webhooks that allow integration with GitHub’s issues tracker~\cite{webhookDoc}. For data storage, \toolname uses the relational database system, SQLite3. Currently, \toolname is running on a modest server and can support five to six repositories concurrently.  For production, it needs to be deployed in a robust server, possibly in a cloud infrastructure. Though \toolname is tailored for handling GitHub issue management, most of its backend components can be easily adapted for other issue trackers.
\looseness=-1


\section{\toolname's Evaluation}
We conducted a preliminary evaluation to measure \toolname's models' predictive performance as well as \toolname's usability and usefulness. \toolname's GitHub repository provides details of the evaluation methodology, results, and necessary data to replicate and verify the evaluation~\cite{repl_pack}. 

\subsection{Model Evaluation}
To select appropriate models for our features, we conducted an extensive literature review, experimented with various models, and chose those that displayed the best predictive performance. Our approach was to replicate the original evaluation of the models by following the methodologies and test datasets provided in their respective papers. For the similar issue detection feature, we chose the RTA model~\cite{fang2023representthemall} fine-tuned in RTA's duplicate issues training dataset~\cite{representThemAllDataset}. Their test dataset~\cite{representThemAllDataset} has 15,288 issue pairs (approximately 60\% of the pairs were duplicates, and the remaining were non-duplicates) from 6 large open-source projects. Overall, we obtained 97.3\% accuracy, 97.5\% precision, and 98.8\% recall. For issue severity prediction, we selected the RTA model~\cite{fang2023representthemall} fine-tuned in RTA's issue severity training dataset~\cite{representThemAllDataset}. For this task, the test dataset has 15,510 issues of 6 projects with 5 severity classes. The distribution of the severity classes was between 15\% and 24\%.  Overall, we obtained 65.6\% accuracy, 67.3\% precision, and 64.6\% recall. For the bug localization feature, we fine-tuned the Llama-2-7b-chat model on the dataset provided by Bogomolov \etal~\cite{longCodeArena}, and we evaluated the model’s prediction capability on their 150 test issues~\cite{longCodeBLDataset}. We achieved 34\% accuracy, 20\% Recall@2 (R@2), 31\% Precision@2 (P@2), and a MAP of 29\%. 
\looseness=-1

\subsection{User Study}
We conducted a user study involving five professional developers with three to eight years of issue management experience who work on different projects at Samsung Research Bangladesh. 
The goal of the study was to evaluate (i) \toolname's usefulness/usability {(RQ1/RQ2)} and (ii) the predictive accuracy of \toolname's suggestions {(RQ3)}.

\subsubsection{Methodology}
For the similar issue detection feature, we first selected 2 issues as queries and their respective duplicates and 8 non-duplicate issues from the RTA~\cite{fang2023representthemall} model's duplicate bug report test dataset~\cite{representThemAllDataset}. These issues come from the OpenOffice project~\cite{openoffice}. To evaluate the severity prediction feature, focusing on the same project, we first categorized the RTA~\cite{fang2023representthemall} model's issue severity test dataset~\cite{representThemAllDataset} into 2 groups: (i) issues for which the model predicted the severity correctly and (ii) issues where the model mispredicted the severity. Then, we chose one issue from each group randomly. For the bug localization feature, we chose one of the projects from Bogomolov \etal's dataset~\cite{longCodeBLDataset}: wso2/testgrid~\cite{wso2}. Each issue in this project had two buggy code files on average. Then, we ran the bug localization model on each issue three times since LLM outputs can vary slightly each time and selected the common suggestions that occurred in all three runs. After that, we divided these issues into two groups: (i) issues for which the model predicted at least one correct buggy code file (\ie present in the ground truth) in its top-5 suggestions and (ii) issues where the model failed to predict any buggy code files in the top-5 suggestions. We selected one issue from each group as queries. Our goal in selecting both successful and unsuccessful cases across various features was to ensure a fair evaluation, allowing participants to experience these scenarios and better assess the tool's usefulness and usability.


In the study, the participants first were introduced to the tool with detailed guidelines. Then, they engaged in answering a questionnaire that assessed the accuracy, usefulness, and usability of the tool's suggestions. The study survey questions were a combination of Likert-scale and open-ended questions. We provided participants with ground truth data for reference.

\subsubsection{RQ1/RQ2 Results}
The participants (\aka users) evaluated how easy-to-use and practical \toolname is.

\underline{Readability of \toolname's Suggestions:} Of five users, three rated \toolname's suggestions for the three features as ``very easy" to understand, while two rated them as ``moderately easy" (on a 5-point Likert scale, these options are the most positive).
\looseness=-1

\underline{{\toolname Overall Usability:}} Four users found \toolname easy to use, while one user was unsure. All the users agreed that \toolname's feedback comments were easy to understand and displayed useful information.

\underline{{\toolname Responsiveness:}} \toolname's average response time for showing results for all three features was approximately 90 seconds (for a repository with $\approx$20 issues and $\approx$50 code files). Four users found \toolname ``very responsive'' and one user found it ``moderately responsive'' (on a 5-point Likert scale, these are the most positive options).

\underline{{\toolname Usefulness for Issue Management:}} Four users ``completely agreed," while one user ``somewhat agreed" with the statement: ``The combination of \toolname's three features is helpful for issue management".

\underline{New feature suggestions}: The participants suggested additional features for \toolname, such as automated issue content identification (\eg identifying the reproduction steps), quality assessment, automated program repair, and developer recommendations for issue resolution.

\subsubsection{RQ3 Results} 
The participants evaluated the accuracy of the suggestions made by the \toolname's three main features.

\underline{{Similar Issue Identification:}} All the users agreed that \toolname correctly suggests all the similar issues for the two queries. However, the reporters found that \toolname suggests one or two extra non-duplicates for the first query---though, it perfectly predicts the two expected similar issues for the second query.

\underline{{Severity Prediction:}} Three of five users agreed that \toolname correctly predicted the severity of the issues, whereas two users were unsure about the predictions. They stated that \toolname correctly identified the severity class for one query and mispredicted the other, but the misprediction was close to the actual severity class (\toolname predicted `Trivial' instead of `Minor').
\looseness=-1

\underline{{Bug Localization:}} 
\rev{All five users identified correct and incorrect predictions, aided by the ground truth references provided in the study survey. Two users identified one correct prediction in the top-5 suggestions sufficient, while three others suggested improving \toolname's accuracy. Overall, all participants valued the feature's responsiveness and potential usefulness.}

\section{Related Work}
\label{sec:related_work}

Existing tools support individual issue management tasks. Find Duplicates~\cite{find_duplicates}, Probot~\cite{probot}, NextBug~\cite{rocha2015nextbug} are plugins for GitHub, Jira~\cite{jira}, and Bugzilla~\cite{bugzilla} that identify related issues. These tools rely on information retrieval or classical machine learning models that process issue text to determine issue similarity. Priority Scheduler~\cite{priorityScheduler} is a Jira~\cite{jira} plugin that assigns priority based on project deadlines. \rev{BugLocalizer~\cite{thung2014buglocalizer} is a Bugzilla~\cite{bugzilla} extension that analyzes bug reports and source code similarities to identify buggy files.} PR-Agent~\cite{pr-agent} is a multi-featured paid tool that applies LLMs, \eg\ ChatGPT~\cite{chatgpt}, to perform tasks such as pull request change classification, automatic code review, and documentation. \rev{Additionally, there are tools for reporting~\cite{song2023burt,song2022toward}, identifying bug report components~\cite{song2020bee}, and assessing bug reproduction steps~\cite{mahmud:icpc2025,chaparro2019assessing}}. Researchers have proposed automated techniques for duplicate issue detection~\cite{fang2023representthemall,zhang2023duplicate,rodrigues2020soft}, severity prediction~\cite{fang2023representthemall,kim2022bug,ali2024bert}, bug localization~\cite{longCodeArena,mahmud2024using}, issue categorization~\cite{somasundaram2012automatic,catolino2019not}, and other issue management tasks~\cite{chaitra2022bug,saha2024toward}. We selected state-of-the-art models, namely RTA~\cite{fang2023representthemall} and LongCodeArena~\cite{longCodeArena} for \toolname's three features, based on a rigorous literature review found in our replication package~\cite{repl_pack}.
\looseness=-1

Compared to prior tools, \toolname stands out as an open-source, easy-to-install solution that seamlessly integrates with GitHub and consolidates multiple features, making it a comprehensive assistant for issue management.
Moreover, \toolname leverages state-of-the-art models that have demonstrated superior performance compared to prior proposed approaches.
\looseness=-1



\section{Conclusions \& Future Work}
\label{sec:conclusions}

\toolname is an integrated open-source GitHub application that leverages state-of-the-art models to identify similar issues, predict issue severity, and localize potential buggy code files for a new issue. 
It aims to support developers in managing and resolving issues.
Evaluation results indicate that \toolname provides valuable assistance in terms of predictive accuracy and user experience. For future work, we plan to enhance the performance of \toolname's bug localization feature by incorporating more advanced models. Additionally, we aim to extend \toolname’s compatibility to platforms beyond GitHub.

\section*{Acknowledgments}
This work is supported by U.S. NSF grants CCF-2239107 and CCF-1955853. The opinions, findings, and conclusions expressed in this paper are those of the authors and do not necessarily reflect the sponsors' opinions.



\balance
\bibliographystyle{IEEEtran}
\bibliography{references}

\end{document}